\documentclass[acmsmall]{acmart}
\usepackage{amsmath,amsfonts}
\usepackage{algorithm}
\usepackage{subfig}
\usepackage{url}
\usepackage[normalem]{ulem}
\usepackage{algpseudocode}

\usepackage{multirow}
\usepackage{xspace}
\newcommand{\modelname}{PTF-FSR\xspace} 
\newcommand{\modelnamenospace}{PTF-FSR}

\AtBeginDocument{%
  }

\setcopyright{acmlicensed}
\copyrightyear{2018}
\acmYear{2018}
\acmDOI{XXXXXXX.XXXXXXX}

\acmJournal{JACM}
\acmVolume{37}
\acmNumber{4}
\acmArticle{111}
\acmMonth{8}

\begin{document}

\title{\modelnamenospace: A Parameter Transmission-Free Federated Sequential Recommender System}

\author{Wei Yuan}
\authornote{Both authors contributed equally to this research.}
\email{w.yuan@uq.edu.au}
\affiliation{%
  \institution{The University of Queensland}
  \city{Brisbane}
  \state{QLD}
  \country{Australia}
}

\author{Chaoqun Yang}
\authornotemark[1]
\email{chaoqun.yang@griffith.edu.au}
\affiliation{%
  \institution{Griffith University}
  \city{Gold Coast}
  \state{QLD}
  \country{Australia}
}

\author{Liang Qu}
\email{liang.qu@uq.edu.au}
\affiliation{%
  \institution{The University of Queensland}
  \city{Brisbane}
  \state{QLD}
  \country{Australia}
}

\author{Quoc Viet Hung Nguyen}
\email{henry.nguyen@griffith.edu.au}
\affiliation{%
  \institution{Griffith University}
  \city{Gold Coast}
  \state{QLD}
  \country{Australia}
}

\author{Guanhua Ye}
\email{g.ye@bupt.edu.cn}
\affiliation{%
 \institution{Beijing University of Posts and Telecommunications}
 \city{Beijing}
 \state{Beijing}
 \country{China}}

\author{Hongzhi Yin}\authornote{Corresponding author.}
\email{h.yin1@uq.edu.au}
\affiliation{%
  \institution{The University of Queensland}
  \city{Brisbane}
  \state{QLD}
  \country{Australia}
}

\renewcommand{\shortauthors}{Yuan et al.}

\begin{abstract}
Sequential recommender systems, as a specialized branch of recommender systems that can capture users' dynamic preferences for more accurate and timely recommendations,
have made significant progress.
Recently, due to increasing concerns about user data privacy, some researchers have implemented federated learning for sequential recommendation, a.k.a., Federated Sequential Recommender Systems (FedSeqRecs), in which a public sequential recommender model is shared and frequently transmitted between a central server and clients to achieve collaborative learning.
Although these solutions mitigate user privacy to some extent, they present two significant limitations that affect their practical usability:
(1) They require a globally shared sequential recommendation model. However, in real-world scenarios, the recommendation model constitutes a critical intellectual property for platform and service providers. Therefore, service providers may be reluctant to disclose their meticulously developed models.
(2) The communication costs are high as they correlate with the number of model parameters.  
This becomes particularly problematic as the current FedSeqRec will be inapplicable when sequential recommendation marches into a large language model era.

To overcome the above challenges, this paper proposes a parameter transmission-free federated sequential recommendation framework (\modelname), which ensures both model and data privacy protection to meet the privacy needs of service providers and system users alike.
Furthermore, since \modelname only transmits prediction results under privacy protection, which are independent of model sizes,  this new federated learning architecture can accommodate more complex and larger sequential recommendation models.
Extensive experiments conducted on three widely used recommendation datasets, employing various sequential recommendation models from both ID-based and ID-free paradigms, demonstrate the effectiveness and generalization capability of our proposed framework.
\end{abstract}



\received{20 February 2007}
\received[revised]{12 March 2009}
\received[accepted]{5 June 2009}

\maketitle

\section{Introduction}\label{sec:introduction}
Recommender systems have served as an integral part of most web services, due to their success in alleviating information overload~\cite{yin2016adapting}.
The heart of recommender systems is to accurately characterize users' personal preferences.
Based on the observation that users' interests are evolved and influenced by their recent historical behaviors in many real-world cases (e.g., e-commerce~\cite{schafer2001commerce}, online news~\cite{wu2020mind} and social media~\cite{yin2016spatio,nguyen2017retaining}), a new branch of recommender systems, named sequential recommender systems, have emerged to discover users' sequential and dynamic behavior patterns~\cite{wang2016spore}. Sequential recommendation models are developed by the platforms or service providers and trained with enormous user-interacted item sequences on a central server. Unfortunately, this centralized training paradigm has been criticized for taking significant risks in private data leakage~\cite{batmaz2019review}.
Given the growing awareness of user privacy and the recent release of privacy protection regulations like GDPR~\cite{voigt2017eu} and CCPA~\cite{harding2019understanding}, it has become increasingly challenging for online platforms to train a sequential recommender using the traditional centralized training paradigm without violating these regulations.

\begin{figure}[!ht]
  \centering
  \includegraphics[width=0.8\textwidth]{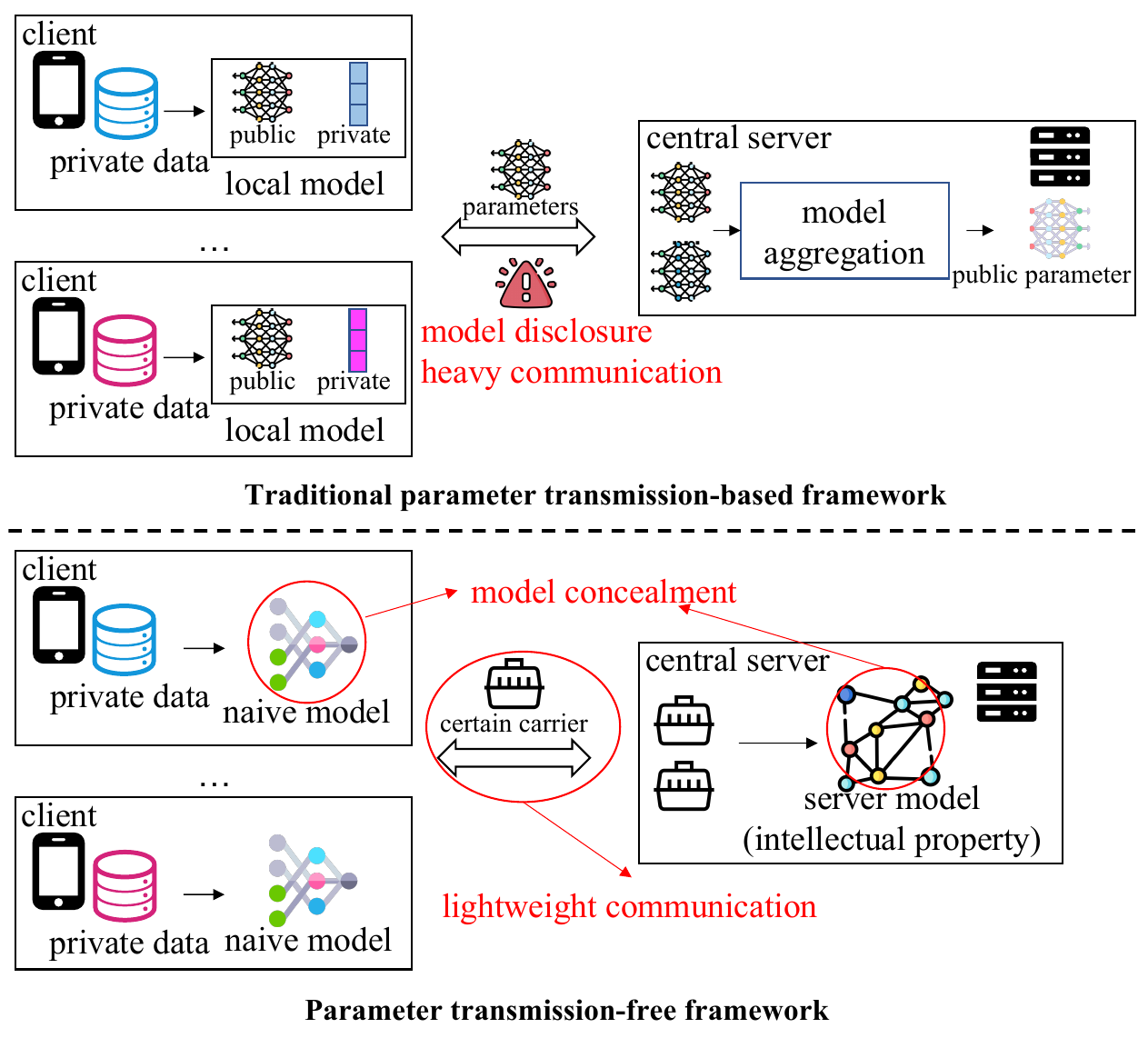}
  \caption{Traditional parameter transmission-based framework v.s. our parameter transmission-free framework. Traditional framework leverages model parameters to transfer knowledge between clients and the central server, therefore, it suffers model disclosure and heavy communication problems. However, our parameter transmission-free framework replaces the parameters with certain carriers, there, our framework can conceal the model and if the carrier is lightweight, the communication cost will be affordable.}\label{fig_difference}
\end{figure}

To address privacy concerns, several studies~\cite{lin2022generic,li2023distvae} have explored adopting the federated learning framework to sequential recommender systems to enhance user privacy, known as FedSeqRecs. In this privacy-conscious training scheme, clients/users\footnote{In this paper, the terms ``client" and ``user" are used interchangeably, as each client is responsible for one user.} can collaboratively develop a model without exposing their private data. 
This is achieved by frequently exchanging locally tuned model parameters between clients and a central server. However, as highlighted in our previous work~\cite{yuan2023hide}, this parameter-transmission-based federated learning framework has two significant drawbacks that restrict its practical usability. We argue that these limitations persist and are even more pronounced in the context of FedSeqRecs.

The first defect of such a framework is that it requires the service provider to open-source its sequential recommendation model to all participants. In a real commercial environment, the sequential model is the core intellectual property of the service providers or online platforms. Sharing the model with clients exposes it to the risk of being stolen by competitors, leading to a significant loss of commercial value derived from the company's technical advantages.
Although some studies in the realm of federated learning have explored the use of digital watermarking to protect intellectual property~\cite{tekgul2021waffle}, digital watermarking only offers limited and unreliable tracking of model copying behavior and lacks the capability to prevent plagiarism. Therefore, the primary strategy for service providers to protect their valuable models is to keep them as commercial secrets. Consequently, current parameter transmission-based FedSeqRecs have limited applicability, as they neglect the model privacy protection needs of service providers and even compromise the platform’s model privacy to protect user privacy.

Another shortcoming of the parameter transmission-based FedSeqRec is that its communication expenses are proportional to the number of sequential model parameters. In sequential recommendation, this setting results in substantial communication costs. Specifically, sequential recommendation models can generally be classified into ID-based and ID-free paradigms. ID-based models, such as GRU4Rec~\cite{jannach2017recurrent} and SASRec~\cite{kang2018self}, have a significant number of parameters dominated by the item embedding table. This table is typically a high-dimensional matrix due to the large number of items. Conversely, the size of ID-free sequential recommendation models does not increase with the number of items. However, these models often rely on the capabilities of large language models~\cite{zhao2024recommender}, which inherently contain a vast number of parameters. For instance, even a BERT-small~\cite{devlin2019bert} version of MoRec~\cite{yuan2023go} incurs a one-time transmission cost of over $100$MB from one client to the central server. Consequently, as model sizes increase, the communication burden becomes impractical for parameter transmission-based FedSeqRec in real-world applications. The upper part of Fig.~\ref{fig_difference} summarizes these two limitations discussed above.

Clearly, both of these limitations stem from the approach of using model parameters to transfer knowledge in FedSeqRecs. Consequently, a natural solution to these issues is to replace the model parameters with more efficient and model-agnostic carriers to convey knowledge between clients and the server. As illustrated in the lower part of Fig.~\ref{fig_difference}, once the carriers are decoupled from the model parameters, the service provider can deploy its sophisticated and confidential sequential recommender model on the server side while implementing simpler and publicly available recommendation models on the client side, thereby protecting the valuable model. Additionally, if the carriers are lightweight, this also reduces communication costs.

In our previous work~\cite{yuan2023hide}, we took the initial step in developing a parameter transmission-free federated recommender framework (PTF-FedRec) using a triple consisting of user, item, and prediction scores $(u_{i}, v_{j}, \hat{r}_{ij})$ as the knowledge carrier.
In PTF-FedRec, client $u_{i}$ uploads the prediction scores $\{\hat{r}_{ij}\}_{j\in \hat{\mathcal{V}}{i}}$ for a set of items to the central server and employs a sampling and swapping mechanism among these predictions to protect user privacy. 
Meanwhile, the central server returns prediction scores for certain items based on their confidence and importance to the client, facilitating knowledge flow between clients and the server.
However, PTF-FedRec is designed for collaborative filtering-based recommendation models and cannot apply to sequential recommenders due to significant differences in both data structures and model architectures. 
Specifically, in the sequential recommendation, the client's data is a sequence $[v_{1}, v_{2}, \dots, v_{n}]$, and the model is trained to predict the next item based on this sequence.
Consequently, the carrier between clients and the central server in sequential recommendation tasks are item sequences. 
As a result, the privacy protection and knowledge transfer mechanisms tailored for triple carriers are not suitable for sequential carriers.

Building on the insights from our previous work~\cite{yuan2023hide}, this paper extends the concept of parameter transmission-free federated recommendation to sequential recommendation, namely \modelname, by incorporating specific designs for sequential privacy preservation and knowledge sharing.
Specifically, in \modelname, the clients and the central server achieve collaborative learning by transmitting sequences.
To ensure user privacy protection, on the client side, we propose an exponential mechanism-based item sequence generation method to add perturbations to users' original item interaction sequences.
In addition, we design two contrastive learning auxiliary tasks, preference consistency and intention similarity contrastive learning, on the central server to mitigate the effects of noise in the client-uploaded sequences by forcing the server model to capture deep and high-level patterns. 
Ultimately, a similarity-based knowledge-sharing method is developed for the central server to sample sequential knowledge for sharing with clients.
To demonstrate the effectiveness of \modelname, we conduct extensive experiments on three popular recommendation datasets using three sequential recommendation models, covering both ID-based (GRU4Rec and SASRec) and ID-free (MoRec) paradigms. 
The experimental results show that \modelname can achieve comparable performance to centralized training methods and significantly outperform existing FedSeqRec baselines in terms of both effectiveness and communication efficiency.

The major new contributions of this paper are listed as follows.
\begin{itemize}
  \item We extend our parameter transmission-free federated recommendation framework proposed in~\cite{yuan2023hide} to the sequential recommendation task by developing a novel parameter transmission-free federated sequential recommendation framework \modelname to solve the intellectual property protection and heavy communication burden problem.
  \item We propose an exponential mechanism-based item sequence generation method to protect user data privacy. Besides, two contrastive learning tasks are designed to facilitate the server model effectively learning from the noisy sequences. Further, a similarity-based knowledge-sharing method is proposed to improve the efficacy of \modelname's collaborative learning.
  \item We conduct extensive experiments on three sequential recommendation datasets with three typical sequential recommender systems, including both ID-based and ID-free recommendation models. The experimental results demonstrate the effectiveness and efficiency of our proposed framework.
\end{itemize}

The remainder of this paper is organized as follows.
The related works of sequential recommender systems, federated recommendations, and intellectual property protection in federated learning are presented in Section~\ref{sec_related_work}.
Section~\ref{sec_preliminary} provides the preliminaries related to our research, including the problem definition of federated sequential recommender systems and the general learning protocol of current federated sequential recommender systems.
Then, in Section~\ref{sec_methodology}, we present the technical details of our proposed \modelname.
The experimental results with comprehensive analysis are exhibited in Section~\ref{sec_experiments}.
Finally, Section~\ref{sec_conclusion} gives a brief conclusion of this paper.

\section{Related Work}~\label{sec_related_work}
In this section, we briefly review the literature on three related topics: sequential recommender systems, federated recommender systems, and model intellectual property protection.
Other involved topics such as the development of general recommender systems and federated learning can be referred to corresponding surveys~\cite{yu2023self,li2021survey}.

\subsection{Sequential Recommender System}
In many online activities, such as online shopping and online reading, users' historical interactions are typically fragmented and contain valuable temporal information. 
Traditional collaborative filtering-based recommender systems~\cite{he2017neural,he2020lightgcn,nguyen2017argument} treat all user-item interactions equally without considering their chronological order, making it difficult to capture users' dynamic and evolving preferences.
To overcome this limitation, Rendle et al.~\cite{rendle2010factorizing} introduced the pioneering sequential recommender system that leverages a first-order Markov chain to model dynamic user preferences.
With the advancement of deep learning, neural networks have been widely applied in sequential recommendation~\cite{fang2020deep}.
For example, Jannach et al.~\cite{jannach2017recurrent} employed gated recurrent units (GRUs) as the backbone of their sequential recommendation model. 
Inspired by the power of feature representation ability, self-attention has been incorporated in models like SASRec~\cite{kang2018self}, which achieved remarkable performance. 
Sun et al.~\cite{sun2019bert4rec} revised the next-item prediction objective function with a Cloze task~\cite{taylor1953cloze} to learn bidirectional sequence transitions.
Recently, the significant achievements of large language models in the NLP~\cite{min2023recent} area have inspired researchers to leverage them in sequential recommendations. 
A new paradigm, ID-free sequential recommendation, has attracted increasing attention.
MoRec~\cite{yuan2023go} was the first to demonstrate the effectiveness of ID-free sequential recommendation by full-finetuning BERT~\cite{devlin2019bert}.
Subsequently, many larger language models have been utilized in sequential recommendation~\cite{zhao2024recommender}.

Although the aforementioned works have achieved significant success, they are trained in a centralized manner, which has been criticized for its privacy risks. In this paper, we select three typical centralized sequential recommendation models (GRU4Rec, SASRec, and MoRec) as our base models and train them using our privacy-preserving framework to demonstrate that \modelname can achieve comparable performance to the centralized methods.

\subsection{Federated Recommender System}
With increasing awareness of privacy protection, the integration of federated learning with recommendation models has emerged as a prominent research topic~\cite{yang2020federated,yin2024device}. 
Ammand et al.~\cite{ammad2019federated} proposed the first federated recommendation framework using collaborative filtering models. 
Subsequently, numerous extensions have been developed to enhance model performance~\cite{lin2020fedrec,qu2023semi,yuan2023hetefedrec}, improve privacy-preserving capabilities~\cite{qu2024towards,zhang2023comprehensive,hung2017computing,yuan2023interaction,yuan2023federated}, and bolster security~\cite{zhang2022pipattack,yuan2023manipulating,yuan2023manipulating1}. 
However, these methods predominantly focus on collaborative filtering-based recommendation models.
FMSS~\cite{lin2022generic} takes the first time to adapt the federated recommendation framework to sequential recommendation tasks.
Li et al.~\cite{li2022federated} explored federated sequential recommendation at the organizational level, while Zhang et al.~\cite{zhang2024feddcsr} investigated FedSeqRec from a cross-domain perspective.

All the aforementioned FedRecs and FedSeqRecs are based on a parameter transmission-based learning protocol. As mentioned in Section~\ref{sec:introduction}, this protocol has limited usability because it overlooks the privacy needs of service providers and incurs substantial communication costs. In our previous work~\cite{yuan2023hide}, we proposed a practical parameter transmission-free framework for collaborative filtering-based recommendations. In this paper, we take a further step by introducing a parameter transmission-free approach to federated sequential recommendation.

\subsection{Contrastive Learning in Recommender System}
Contrastive learning has achieved significant success in recommender systems~\cite{yu2023self,yu2023xsimgcl} by offering self-supervised signals. 
The essential idea of contrastive learning is to minimize the distance between positive instances while pushing the negative instances farther apart in the representation space~\cite{yuan2024robust}. 
S$^{3}$-Rec~\cite{zhou2020s3} devises four auxiliary self-supervised objectives to learn the correlations among attributes, items, subsequences, and sequences. 
CL4SRec~\cite{xie2022contrastive} designs three data augmentation methods, item crop, item mask, and item reorder, to construct different views for contrastive learning in sequential recommendation. 
Additionally, contrastive learning has been used to improve model performance by learning from noisy data. Wu et al.~\cite{wu2022multi} employed multi-view contrastive learning and pseudo-Siamese networks to address the noisy interaction problem. 
KACL~\cite{wang2023knowledge} performs contrastive learning between the knowledge graph view and the user-item interaction graph to eliminate interaction noise. 
He et al.~\cite{he2023robust} leveraged contrastive learning for denoising in the basket recommendation scenario.

In this paper, we design two contrastive learning methods to improve consistency from the perspectives of preference and intention, thereby enabling the server model to learn high-level representations from clients' noisy uploaded sequences.

\subsection{Model Intellectual Property in Federated Learning}
The protection of model intellectual property in federated learning remains an under-explored area. 
Digital watermarking~\cite{tekgul2021waffle,lansari2023federated} is the most commonly used strategy to verify whether a model has been illegally copied and redistributed by adversaries. 
However, designing a durable and accurate watermark without compromising model performance is challenging, especially for recommendation tasks, where users have diverse preferences and items span various categories. 
Moreover, watermarking can only provide verification but does not prevent model plagiarization, which still results in value loss for model owners. 
Therefore, we contend that, as of now, the best way to protect model assets is to keep them undisclosed.

\begin{table}[]
  \centering
  \caption{List of important notations.}\label{tb_notation}
  \begin{tabular}{l|l}
  \hline
   $\mathcal{D}_{u_{i}}$ & the local dataset for user $u_{i}$, which is usually a sequence of interacted items.  \\
   $\hat{\mathcal{D}}_{u_{i}}^{l}$ & the dataset created by user $u_{i}$'s local model in $l$ round. \\
   $\widetilde{\mathcal{D}}_{u_{i}}$ & the dataset created by server model for user $u_{i}$. \\
   $T_{u_{i}}$ & the length of user $u_{i}$'s interaction sequence.  \\
   \hline
   $\mathcal{U}$ & all users in the federated recommender system.  \\
   $\mathcal{U}^{l}$ & selected training users in $l$ round.  \\
   $\mathcal{U}_{u_{i}}^{l}$ & the set of users that interacted similar items with $u_{i}$ in $\mathcal{U}^{l}$.  \\
   $\mathcal{V}$ & all items in the federated recommender system. \\
   $\mathcal{V}_{u_{i}}^{'}$ & user $u_{i}$'s trained items. \\
   \hline
   $r_{ij}$ & the preference score of user $u_{i}$ for item $v_{j}$.   \\ 
   $\hat{r}_{ij}$ & the predicted score for item $v_{j}$ by user $u_{i}$'s local model.   \\ 
   $\widetilde{r}_{ij}^{t}$ & the predicted score of $u_{i}$ for item $v_{j}$ at position $t$ by server model.   \\ 
   \hline
   $\mathbf{M}_{u_i}^{l}$ & user $u_{i}$'s model parameters in round $l$.   \\ 
   $\mathbf{M}_{s}^{l}$ & server model parameters in round $l$.   \\ 
   $\mathbf{e}_{u_{i}}^{t}$ & the latent vector of $u_{i}$ at position $t$ of a sequence.   \\ 
   $\mathcal{F}_{u_{i}}$ & users' model algorithm.   \\ 
   $\mathcal{F}_{s}$ & server model algorithm.   \\ 
   \hline
   $\beta$ & the proportion of items using exponential mechanism-based generation.   \\
   $\epsilon$ & privacy factor in exponential mechanism-based item generation.   \\ 
   $\lambda_{pc}$ & the factor controls the strengths of preference consistency contrastive learning.   \\ 
   $\lambda_{is}$ & the factor controls the strengths of intention similarity contrastive learning.   \\ 
   \hline
  \end{tabular}
  \end{table}

\section{Preliminaries}\label{sec_preliminary}
In this paper, bold lowercase (e.g., $\mathbf{a}$) represents vectors, bold uppercase (e.g., $\mathbf{A}$) indicates matrices, and the squiggle uppercase (e.g., $\mathcal{A}$) denotes sets or functions.
The important notations are listed in Table~\ref{tb_notation}.
\subsection{Problem Definition of Federated Sequential Recommendation}
Let $\mathcal{U}=\{u_{i}\}_{i=1}^{\left|\mathcal{U}\right|}$ and $\mathcal{V}=\{v_{j}\}_{j=1}^{\left|\mathcal{V}\right|}$ denote all users and items, respectively.
$\left|\mathcal{U}\right|$ and $\left|\mathcal{V}\right|$ are total numbers of clients and items.
In FedSeqRec, a user $u_{i}$ owns its private dataset $\mathcal{D}_{u_{i}}=[v_{1}^{u_{i}}, v_{2}^{u_{i}}, \dots, v_{t}^{u_{i}}, \dots, v_{T_{u_{i}}}^{u_{i}}]$, which is a chronological (i.e., $ 1\le t \le T_{u_{i}}$) item interaction log.
For training purposes, $\mathcal{D}_{u_{i}}$ also includes negative samples at each time step, which is usually randomly sampled from a non-interacted item pool.
Note that to ensure user privacy, $\mathcal{D}_{u_{i}}$ is stored in $u_{i}$'s local device and all other participants cannot access it.
The goal of FedSeqRec is to train a sequential recommendation model that can predict user's potentially preferred items by ranking the prediction scores $\hat{r}_{ij}^{T_{u} + 1}$ at time step $T_{u} + 1$.

\subsection{Traditional Parameter Transmission-based Federated Sequential Recommendation Framework}\label{sec_general_fedrec}
To achieve the above goal, existing FedSeqRec systems typically utilize a federated learning protocol that involves parameter transmission coordinated by a central server.
Initially, the central server sets up a sequential recommendation model. If this model includes user embeddings, these parameters are treated as private and are initialized by each client individually.
After that, the central server and clients are coordinated to repeatedly execute the following steps until model convergence.
Firstly, the central server selects a group of clients for training and disperses the sequential recommendation models to them.
Then, the selected clients train the received sequential model on their local datasets $\mathcal{D}_{u_{i}}$ with a specific objective function $\mathcal{L}^{rec}$, for example:
\begin{equation}\label{eq_rec_loss}
  \mathcal{L}^{rec} = -\sum\limits_{v_{j} \in \mathcal{D}_{u_{i}}=[v_{1}^{u_{i}}, v_{2}^{u_{i}}, v_{t}^{u_{i}}, \dots, v_{T_{u_{i}}}^{u_{i}}]} \left[log(\sigma(\hat{r}_{ij})) + \sum\limits_{v_{k}\notin \mathcal{D}_{u_{i}}} log(1 - \sigma(\hat{r}_{ik}))\right]
\end{equation}
Once the local training is complete, the clients upload their trained models back to the central server. The central server then employs a strategy, such as FedAvg~\cite{mcmahan2017communication}, to aggregate these models.
Algorithm~\ref{alg_traditional} summarizes the traditional federated sequential recommendation framework with pseudo-code.

\begin{algorithm}[!ht]
  \renewcommand{\algorithmicrequire}{\textbf{Input:}}
  \renewcommand{\algorithmicensure}{\textbf{Output:}}
  \caption{The pseudo-code for traditional parameter transmission-based federated sequential recommendation.} \label{alg_traditional}
  \begin{algorithmic}[1]
    \Require global round $L$; learning rate $lr$, \dots
    \Ensure  well-trained sequential model $\mathbf{M}^{L}$
    \State server initializes model $\mathbf{M}^{0}$
    \For {each round l =0, ..., $L-1$}
      \State sample a fraction of clients $\mathcal{U}^{l}$ from $\mathcal{U}$
        \For{$u_{i}\in \mathcal{U}^{l}$ \textbf{in parallel}} 
        \State // execute on client sides
        \State $\mathbf{M}_{u_{i}}^{l+1}\leftarrow$\Call{ClientTrain}{$u_{i}$, $\mathbf{M}^{l}$}
        \EndFor
      \State // execute on central server
      \State $\mathbf{M}^{l+1}\leftarrow$ aggregate received client model parameters $\{\mathbf{M}_{u_{i}}^{l+1}\}_{u_{i}\in \mathcal{U}^{l}}$
    \EndFor
    \Function{ClientTrain} {$u_{i}$, $\mathbf{M}^{l}$}
    \State $\mathbf{M}_{u_i}^{l+1}\leftarrow$ update local model with recommendation objective $\mathcal{L}^{rec}$
    \State \Return $\mathbf{M}_{u_{i}}^{l+1}$
    \EndFunction
    \end{algorithmic}
\end{algorithm}

\section{Methodology}\label{sec_methodology}
\begin{figure*}[!htbp]
  \centering
  \includegraphics[width=1.\textwidth]{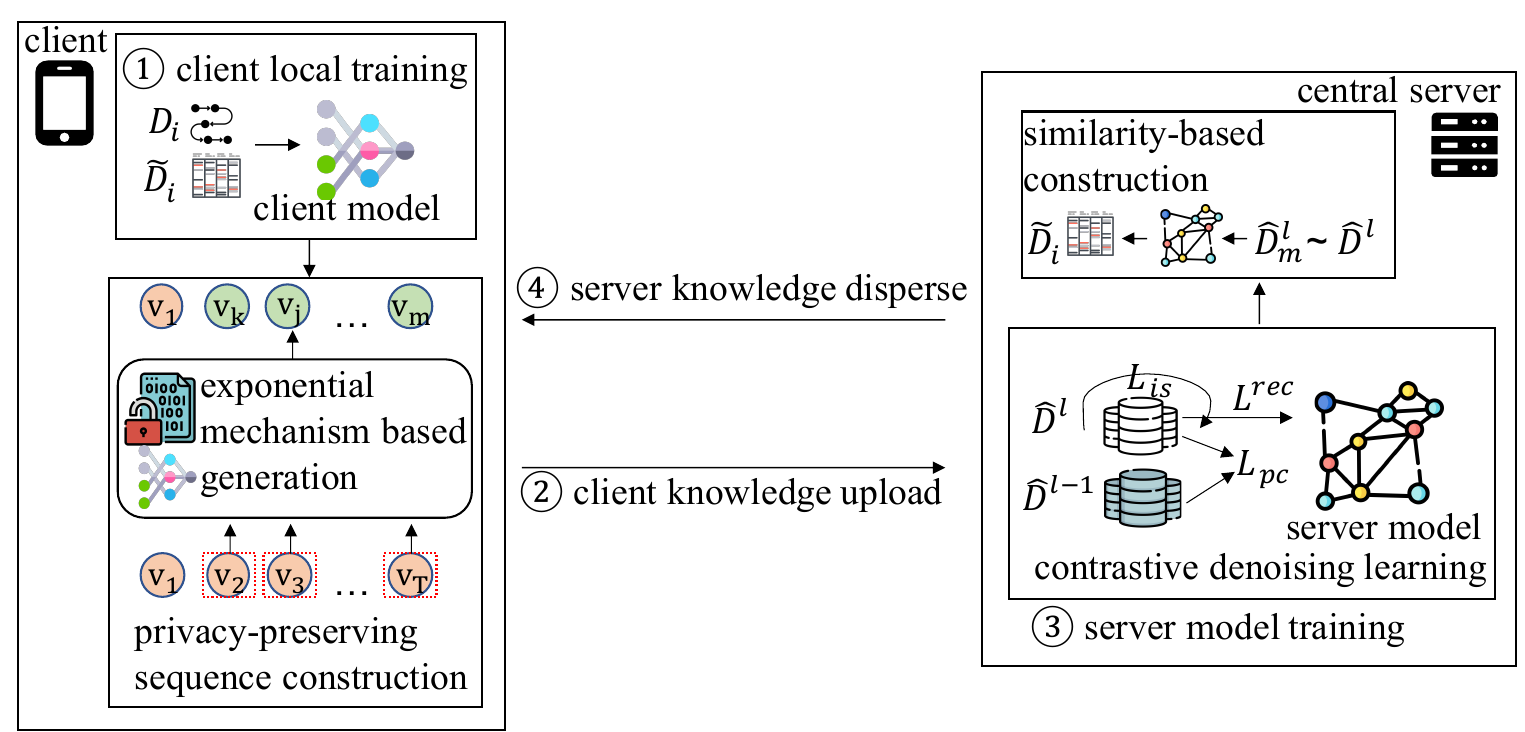}
  \caption{\modelname includes four steps. Clients first train their client models on local datasets. After that, they utilize the trained client models to generate sequences with the exponential mechanism and send the sequences to the central server. The server trains its delicate model on the noisy data with several contrastive auxiliary tasks. Finally, the central server utilizes the trained server model to return some knowledge back to clients.}\label{fig_our_framework}
\end{figure*}

In this section, we first provide a brief introduction to the basic sequential recommendation models utilized in our framework in Section~\ref{sec_basic_srm}.
We then detail the technical aspects of \modelname, which is designed to protect both user and model privacy while also minimizing communication costs.
Specifically, we introduce the overall learning protocol of \modelname in Section~\ref{sec_ptf_protocal}, and then, we present the privacy-preserving client knowledge uploading in Section~\ref{sec_client_privacy_preserving}.
In Section~\ref{sec_server_advanced_training}, we describe the advanced server model training with two contrastive learning tasks while Section~\ref{sec_server_advanced_knowledge} shows similar knowledge sharing from the server to the client side.
An overview of \modelname is illustrated in Fig.~\ref{fig_our_framework}. Additionally, Algorithm~\ref{alg_ours} presents the pseudo-code for \modelname.

\subsection{Basic Sequential Recommendation Models}\label{sec_basic_srm}
Generally, sequential recommendation models can be divided into two categories based on the item embedding methods used: ID-based sequential recommendation and ID-free sequential recommendation. 
Given that a practical federated sequential recommendation framework should be model-agnostic and compatible with most sequential recommender systems, this paper selects three typical sequential recommendation models (GRU4Rec, SASRec, and MoRec) covering both paradigms as our basic sequential models to demonstrate the generalization of our framework.
In the following sections, we will introduce these two paradigms along with the three specific sequential models.

\textbf{ID-based Paradigm.}
The general workflow of ID-based sequential recommendation can be summarized as follows.
Given a sequence of interacted items $[v_{1}, v_{2}, \dots, v_{n}]$ as the input, the ID-based sequential recommendation model first converts them into a sequence of embedding vectors $[\mathbf{v}_{1}, \mathbf{v}_{2}, \dots, \mathbf{v}_{n}]$ using a $\left|\mathcal{V}\right| \times d_{v}$ item embedding table $\mathbf{V}$.
$d_{v}$ is the item embedding dimension.
Subsequently, a specific sequential neural network model $\mathcal{F}$ is applied to the embedding sequences, transforming them into a sequence of latent vectors $[\mathbf{e}_{1}, \mathbf{e}_{2}, \dots, \mathbf{e}_{n}]$.
Based on these latent vectors, the sequential recommendation model calculates prediction score $\hat{r}_{j}$ and orders a recommendation list by the scores.

For instance, in GRU4Rec~\cite{jannach2017recurrent} and SASRec~\cite{wu2021self}, $\mathcal{F}$ is instantiated using GRU~\cite{chung2014empirical} and Transformer~\cite{vaswani2017attention} respectively.
These models leverage the production of latent vectors with item embedding tables to compute item prediction scores for user $u_{i}$ at time step $t+1$:
\begin{equation}\label{eq_original_predict_score}
  \hat{\mathbf{r}}^{t+1}_{i*} = \mathbf{e}^{t\top}_{u_{i}} \mathbf{V}
\end{equation}

\textbf{ID-free Paradigm.}
Attracting by the powerful representation capabilities of large language models~\cite{min2023recent}, some recent sequential recommendation systems have begun replacing traditional item embedding tables with language model encoders, denoted as $\mathcal{F}_{lm}$.

For example, in MoRec~\cite{yuan2023go}, $\mathcal{F}_{lm}$ utilizes BERT~\cite{devlin2019bert} to process metadata (e.g., titles) of an item $v_{j}$ to generate its embedding $\mathbf{v}_{j}$ rather than using an ID-based embedding table.
After that, the item prediction calculation process is similar to that of the ID-based paradigm.
It is worth noting that, unlike traditional works~\cite{zheng2017joint} that freeze $\mathcal{F}_{lm}$ and only use the large language model as text encoder to extract the textual feature as side information, $\mathcal{F}_{lm}$ in MoRec is full-finetuning and the textual feature is deemed as the main feature of the item.
As a result, the amount of trained parameters in MoRec is much larger than in the traditional sequential recommendation model, and the parameters transmission-based FedSeqRec cannot afford its training process.
By contrast, as will be shown in the remaining paper, the communication cost of our \modelname is model-agnostic, thus, \modelname is compatible with the large language model-based MoRec.

\begin{algorithm}[!ht]
  \renewcommand{\algorithmicrequire}{\textbf{Input:}}
  \renewcommand{\algorithmicensure}{\textbf{Output:}}
  \caption{The pseudo-code for \modelname.} \label{alg_ours}
  \begin{algorithmic}[1]
    \Require global round $L$; learning rate $lr$, \dots
    \Ensure  well-trained server model $\mathbf{M}_{s}^{L}$
    \State server initializes model $\mathbf{M}_{s}^{0}$, clients initialize $\mathbf{M}_{u_i}^{0}$
    \State $\{\widetilde{\mathcal{D}}_{u_{i}}=\emptyset\}_{u_{i}\in\mathcal{U}}$
    \For {each round l =0, ..., $L-1$}
      \State sample a fraction of clients $\mathcal{U}^{l}$ from $\mathcal{U}$
        \For{$u_{i}\in \mathcal{U}^{l}$ \textbf{in parallel}} 
        \State // execute on client sides
        \State $\hat{\mathcal{D}}_{u_{i}}^{l}\leftarrow$\Call{ClientTrain}{$u_{i}$, $\widetilde{\mathcal{D}}_{u_{i}}$}
        \EndFor
      \State // execute on central server
      \State receive client prediction datasets $\{\hat{\mathcal{D}}_{u_{i}}^{l}\}_{u_{i}\in\mathcal{U}^{l}}$
      \State $\mathbf{M}_{s}^{l+1}\leftarrow$ update server model according to Section~\ref{sec_server_advanced_training}
      \State update $\{\widetilde{\mathcal{D}}_{u_{i}}\}_{u_{i}\in\mathcal{U}^{l}}$ according to Section~\ref{sec_server_advanced_knowledge}
    \EndFor
    \Function{ClientTrain} {$u_{i}$, $\widetilde{\mathcal{D}}_{u_i}$}
    \State $\mathbf{M}_{u_i}^{l+1}\leftarrow$ update local model using E.q.~\ref{eq_client_objective}
    \State construct $\hat{\mathcal{D}}_{u_i}^{l}$ according to Section~\ref{sec_client_privacy_preserving}
    \State \Return $\hat{\mathcal{D}}_{u_i}^{l}$
    \EndFunction
    \end{algorithmic}
\end{algorithm}

\subsection{The Parameter Transmission-free Sequential Recommendation Learning Protocol} \label{sec_ptf_protocal}
The fundamental concept of our learning protocol is to utilize sequences generated by clients and the central server to facilitate knowledge transfer between both parties. 
In this arrangement, the service provider does not need to expose its valuable model, ensuring that model privacy requirements are met. 
Moreover, if sequence sharing incorporates privacy-preserving measures, it also safeguards user privacy. 
Consequently, the protocol can successfully preserve both model and user privacy. 
Additionally, the costs associated with transmitting sequences are significantly lower than those for sending model parameters and are independent of model sizes. 
Therefore, the communication overhead in such a learning protocol is considerably reduced.
In this subsection, we first present the general learning protocol of our framework, and leave the introduction of specific privacy-preserving and learning mechanisms in the following subsections. 

\textbf{Initial Stage.}
The central server initializes an elaborately designated sequential recommendation model $\mathcal{F}_{s}$ 
with parameters $\mathbf{M}_{s}^{0}$, while the clients $u_{i}$ initialize some simple and publicly available sequential recommendation models $\mathcal{F}_{u_{i}}$ with parameters $\mathbf{M}_{u_{i}}^{0}$.
Subsequently, the clients and central server iteratively execute the following steps to achieve collaborative learning.

\textbf{Client Model Training.}
At the global round $l$, a group of clients $\mathcal{U}^{l}$ are selected to join the collaborative learning process.
The clients will train their local sequential recommendation model on their local datasets.
In \modelname, clients' local datasets contain two parts: their corresponding private data $\mathcal{D}_{u_{i}}$ and the augmented dataset $\widetilde{\mathcal{D}}_{u_{i}}$ received from the central server.
When training on $\mathcal{D}_{u_{i}}$, following most sequential recommendation settings~\cite{jannach2017recurrent,wu2021self,yuan2023go}, for each time step, we randomly sample some non-interacted items as negative samples.
While for $\widetilde{\mathcal{D}}_{u_{i}}$, each sequence already has a group of items' with soft labels at each time step, i.e., $\widetilde{\mathcal{D}}_{u_{i}}= \{\widetilde{S}_{u_{i},k}\}_{k=1}^{\left|\widetilde{\mathcal{D}}_{u_{i}}\right|}$ and $\widetilde{S}_{u_{i},k}=[\{(v_{j}, \widetilde{r}_{ij}^{1})\}_{j\in \mathcal{V}_{1}}, \dots, \{(v_{k}, \widetilde{r}_{ik}^{t})\}_{j\in \mathcal{V}_{t}},\dots]$.
$\left|\widetilde{\mathcal{D}}_{u_{i}}\right|$ is the number of sequences that are received from the central server.
The model is trained to match the soft labels.
Formally, the local training process can be described as follows:
\begin{equation}
  \label{eq_client_objective}
  \begin{aligned}
    &\mathbf{M}_{u_{i}}^{l+1} = \mathop{argmin}\limits_{\mathbf{M}_{u_{i}}^{l}} \mathcal{L}^{c}(\mathcal{F}_{u_{i}}(\mathbf{M}_{u_{i}}^{l})|\mathcal{D}_{u_{i}}\cup\widetilde{\mathcal{D}}_{u_{i}})\\
  \end{aligned}
\end{equation}
where $\mathcal{L}^{c}$ is the recommendation loss function $\mathcal{L}^{rec}$.

\textbf{Client Knowledge Uploading.}
After local training, clients will transfer their knowledge to the central server.
In \modelname, the clients achieve knowledge sharing by sending sequences $\hat{\mathcal{D}}_{u_{i}}^{l}$ to the central server.
In Section~\ref{sec_client_privacy_preserving}, we will discuss how to construct $\hat{\mathcal{D}}_{u_{i}}^{l}$ considering both data utility and user privacy protection.

\textbf{Server Model Training.}
The central server receives sequences from clients $\{\hat{\mathcal{D}}_{u_{i}}^{l}\}_{u_{i}\in \mathcal{U}^{l}}$ at round $l$ and trains its valuable sequential model on these data:
\begin{equation}
  \label{eq_server_naive_objective}
  \begin{aligned}
    \mathbf{M}_{s}^{l+1} = \mathop{argmin}\limits_{\mathbf{M}_{s}^{l}}\sum\limits_{u_{i}\in \mathcal{U}^{l}} \mathcal{L}^{s}(\mathcal{F}_{s}(\mathbf{M}_{s}^{l})|\hat{\mathcal{D}}_{u_{i}}^{l})\\
  \end{aligned}
\end{equation}
where $\mathcal{L}^{s}$ indicates the objective function used on the central server.
Naively, $\mathcal{L}^{s}$ can be the recommendation objective $\mathcal{L}^{rec}$, but it can achieve limited performance since $\{\hat{\mathcal{D}}_{u_{i}}^{l}\}_{u_{i}\in \mathcal{U}^{l}}$  usually contains many noise in order to protect user privacy.
In Section~\ref{sec_server_advanced_training}, we will discuss how to fully mine the noisy dataset $\{\hat{\mathcal{D}}_{u_{i}}^{l}\}_{u_{i}\in \mathcal{U}^{l}}$ by adding two auxiliary contrastive learning tasks.

\textbf{Server Knowledge Sharing.}
Since the server's model is trained on massive clients' uploaded sequences, it will achieve a more powerful recommender model.
Therefore, after server model training, the central server will disperse knowledge back to promote clients' local training, so that clients can upload higher quality sequences and finally train a better server model.
To be specific, for a client $u_{i}$, the central server samples a sequence from its training set $\{\hat{\mathcal{D}}_{u_{i}}^{l}\}_{u_{i}\in \mathcal{U}^{l}}$ and generates prediction scores for items in this sequence served as soft labels.
Formally,
\begin{equation}
  \label{eq_server_naive_sample}
  \begin{aligned}
     \hat{\mathcal{D}}_{u_{m}}^{l} \sim sample(\{\hat{\mathcal{D}}_{u_{j}}^{l}\}_{u_{j}\in \mathcal{U}^{l}})
  \end{aligned}
\end{equation}
\begin{equation}
  \label{eq_server_naive_generate}
  \begin{aligned}
    \{(v_{k}, \widetilde{r}_{ik}^{t})\}_{k\in \mathcal{V}_{t}}=\mathcal{F}_{s}(\mathbf{M}_{s}^{l+1}, \hat{\mathcal{D}}_{u_{m}}^{l}[1:t-1], \mathcal{V}_{t})
  \end{aligned}
\end{equation}
where $\mathcal{V}_{t}$ includes the original item $v_{t}$ and negative samples at time step $t$.
By calculating E.q.~\ref{eq_server_naive_sample} and \ref{eq_server_naive_generate} several times, the central server constructs $\widetilde{\mathcal{D}}_{u_{i}}$ for client $u_{i}$, which contains $\left|\widetilde{\mathcal{D}}_{u_{i}}\right|$ number of sequences.
In Section~\ref{sec_server_advanced_knowledge}, we will investigate how to sample appropriate sequence $\hat{\mathcal{D}}_{u_{m}}^{l}$ to benefit local user training.

The above is the training protocol of \modelname.
By repeatedly executing the above steps with $L$ rounds, the central server finally obtains a well-trained sequential recommendation model.
Then, during the inference stage, the client queries the central server by sending privacy-preserving $\hat{\mathcal{D}}_{u_{i}}^{L}$ and the server model gives recommendations based on the query.

\subsection{Privacy-preserving Client Sequence Construction}\label{sec_client_privacy_preserving}
The quality of $\hat{\mathcal{D}}_{u_{i}}^{l}$ is crucial for the final performance of the server model in \modelname, and it is also the primary source that may disclose user privacy. 
Therefore, it is essential to design a privacy-preserving mechanism that effectively balances both the utility and privacy protection of $\hat{\mathcal{D}}_{u_{i}}^{l}$.

\textbf{Exponential Mechanism-based Item Generation.}
We at first design an exponential mechanism-based item generation that only considers achieving strict privacy-preserving ability.
Specifically, clients use their trained sequential models to generate new sequences for $\hat{\mathcal{D}}_{u_{i}}^{l}$ based on their original interaction logs:
\begin{equation}
  \begin{aligned}
    \hat{\mathcal{D}}_{u_{i}}^{l} \leftarrow \mathcal{F}_{u_{i}}(\mathbf{M}_{u_{i}}^{l+1}, \mathcal{D}_{u_{i}})
  \end{aligned}
\end{equation}
Note that different from the server model's knowledge sharing (E.q.~\ref{eq_server_naive_generate}), clients only generate sequences without soft labels considering the following reasons: 
(1) Soft labels may contain additional information that could potentially compromise user privacy;
(2) The local client model is trained on limited local data resources, therefore, the soft labels may contain even more noises that impede the server model's training, especially in the initial few rounds.

To protect user privacy, we novelly incorporate exponential mechanism~\cite{mcsherry2007mechanism} during the sequential model's generation process.
Specifically, the possibility of an item $v_{j}$ that will be selected at the $t$'th position of the sequence in $\hat{\mathcal{D}}_{u_{i}}^{l}$ is as follows:
\begin{equation}
  \label{eq_client_predict}
  \begin{aligned}
    Pr[\hat{v}=v_{j}] = \frac{exp(\frac{\epsilon}{2\Delta} r_{ij})}{\sum\limits_{v_k\in \mathcal{V}_{u_{i}}^{'}}exp(\frac{\epsilon}{2\Delta} r_{ik})} 
  \end{aligned}
\end{equation}
where $r_{ij}$ is the prediction score calculated based on previous interacted items (e.g., E.q.~\ref{eq_original_predict_score}) and we omit the time step subscript here to be concise. 
$\mathcal{V}_{u_{i}}^{'}$ is the set of trained items for user $u_{i}$.
$\epsilon>0$ is the privacy budget and $\Delta$ is the sensitivity.
Based on the probability $Pr[\hat{v}=v_{j}]$, we randomly sample the $t$'th item from $\mathcal{V}_{u_{i}}^{'}$ to generate $\hat{\mathcal{D}}_{u_{i}}^{l}$.
As E.q.~\ref{eq_client_predict} fully satisfies the exponential mechanism, generating items at each time step is $\epsilon$-differential private~\cite{mcsherry2007mechanism}.
Thus, according to composition theorem~\cite{dwork2014algorithmic}, the generated sequence will be $\epsilon*T_{u_{i}}$ differential private.

Unfortunately, fully utilizing the above method to generate the entire sequence of $\hat{\mathcal{D}}_{u_{i}}^{l}$ would significantly compromise model performance, even with a very lenient $\epsilon$ privacy budget. 
This is because the entire sequence is sampled with randomness. Although the randomness is based on the expected distribution of the client model prediction, the local client model is far from being well-trained due to limited training resources, especially in the initial few rounds. Therefore, the sequence with full randomness will lose most semantic meanings and disturb the server model training. 

To strike a balance between utility and privacy protection, we propose partially utilizing the exponential mechanism-based item generation method.
Specifically, for the original sequence $\mathcal{D}_{u_{i}}$, we randomly select a ratio $\beta$ of items to replace them with items generated using the exponential mechanism, while keeping the remaining items unchanged.
Since the processed sequence contains a mixture of real interacted items and sampled items, the central server still cannot discern the client's historical interactions, thus preserving user privacy.

\subsection{Contrastive Denoising Learning Mechanism} \label{sec_server_advanced_training}
Effectively utilizing the clients' uploaded datasets $\{\hat{\mathcal{D}}_{u_{i}}^{l}\}_{u_{i}\in \mathcal{U}^{l}}$ is not trivial, as they contain significant noise in each sequence to protect user privacy.
To enable the server model to accurately capture user behavior patterns from these noisy sequences, except for the recommendation task, we design two contrastive learning-based auxiliary tasks that encourage the model to mine deep and high-level sequential knowledge.

\textbf{Preference Consistency Contrastive Learning.}
Although a client $u_{i}$ will upload $\hat{\mathcal{D}}_{u_{i}}^{l}$ with different noise at each round $l$, the underlying preferences within these sequences should be similar, as they originate from the same user.
Building upon this insight, we propose treating the sequences uploaded by the same user at each round as a positive view, aiming to minimize the feature distances between these sequences. 
In addition, considering that the user's model is updated each round, the sequences uploaded from several rounds ago intuitively may be much different from the more recent one, we only apply the preference consistency contrastive learning on the two most recent uploaded datasets $\hat{\mathcal{D}}_{u_{i}}^{l}$ and $\hat{\mathcal{D}}_{u_{i}}^{l-1}$ described as follows:
  \begin{equation}
    \label{eq_preference_cl}
    \begin{aligned}
      \mathcal{L}^{pc}\!=\!-log\frac{exp(sim(\mathbf{e}_{u_{i}}^{l-1}\!,\!\mathbf{e}_{u_{i}}^{l}))}{exp(sim(\mathbf{e}_{u_{i}}^{l-1}\!,\!\mathbf{e}_{u_{i}}^{l})) + \sum\limits_{u_{j}\in \mathcal{U}^{l}/\mathcal{U}^{l}_{u_{i}}}exp(sim(\mathbf{e}_{u_{i}}^{l}\!,\!\mathbf{e}_{u_{j}}^{l})) }\\
    \end{aligned}
  \end{equation}
where $\mathbf{e}_{u_{i}}^{l} \leftarrow F_{s}(\mathbf{M}_{s}^{l}, \hat{\mathcal{D}}_{u_{i}}^{l})$ is the sequence representation from $\hat{\mathcal{D}}_{u_{i}}^{l}$. In this paper, we utilize the preference vector at the last time step (i.e., $T_{u_{i}}$) as the sequence representation.
$sim(x,y)$ is the similarity between $x$ and $y$ and we leverage cosine similarity to calculate it.

\textbf{Intention Similarity Contrastive Learning.}
In addition to denoising the sequence representation by ensuring the consistency of the same user's preference vector, we also calibrate the learned representation from the perspective of intention similarity.
Intuitively, if two users, $u_{i}$ and $u_{j}$, target similar interactions at time step $t+1$, their sequence representations up to $t$ should also be similar. 
Therefore, for each user $u_{i}$, we use the embedding of their final interacted item $v_{T_{u_{i}}}^{u_{i}}$ to identify a group of users, denoted as $\mathcal{U}_{u_{i}}^{l}$, who have interacted with the most similar item in their last time step.
We then aim to maximize the similarity between $u_{i}$ and the users in $\mathcal{U}_{u_{i}}^{l}$, while treating other users in $\mathcal{U}^{l}$ as negative samples:
\begin{equation}
  \label{eq_intent_cl}
    \mathcal{L}^{is}\!=\!-log\frac{\sum\limits_{u_{k}\!\in\!\mathcal{U}_{u_{i}}}\!exp(sim(\mathbf{e}^{l}_{u_{i}}\!,\!\mathbf{e}_{u_{k}}^{l}))}{\sum\limits_{u_{k}\!\in\!\mathcal{U}_{u_{i}}}\!exp(sim(\mathbf{e}^{l}_{u_{i}}\!,\!\mathbf{e}_{u_{k}}^{l}))\!+\!\sum\limits_{u_{j}\!\in\!\mathcal{U}^{l}\!/\!\mathcal{U}^{l}_{u_{i}}}\!exp(sim(\mathbf{e}_{u_{i}}^{l}\!,\!\mathbf{e}_{u_{j}}^{l})) }\\
\end{equation}

As a result, the final learning objective function of the server model in \modelname is as follows:
\begin{equation}
  \label{eq_server_final_objective}
    \mathcal{L}^{s}= \mathcal{L}^{rec} + \lambda_{pc} \mathcal{L}^{pc} + \lambda_{is} \mathcal{L}^{is}
\end{equation}
$\lambda_{pc}$ and $\lambda_{is}$ are factors that control the strengths of $\mathcal{L}^{pc}$ and $\mathcal{L}^{is}$ respectively.

\subsection{Similarity-based Knowledge Downloading}\label{sec_server_advanced_knowledge}
A client model that performs well can enhance the training of the server model, as the latter is trained using the predictions from the former. 
Therefore, in \modelname, after training the server model, the central server sends some sequences with soft labels back to the clients, as shown in Eq.~\ref{eq_server_naive_generate}.
The key challenge lies in selecting the appropriate $\hat{\mathcal{D}}_{u_{m}}^{l}$ (E.q.~\ref{eq_server_naive_sample}).
Similar to the motivation of Intention Similarity Contrastive Learning, training on sequences with similar intentions may help the local model better understand its owner's preferences.
Consequently, for user $u_{i}$, we sample $\hat{\mathcal{D}}_{u_{m}}^{l}$ from the data uploaded by its similar user group $\mathcal{U}^{l}_{u_{i}}$:
\begin{equation}
  \label{eq_server_final_sample}
    \hat{\mathcal{D}}_{u_{m}}^{l} \sim sample(\{\hat{\mathcal{D}}_{u_{j}}^{l}\}_{u_{j}\in \mathcal{U}^{l}_{u_{i}}})
\end{equation} 

\subsection{Discussion}

\subsubsection{Privacy Protection Discussion}
From the service provider's perspective, unlike traditional federated sequential recommendation methods, all information related to the elaborately designed model, including model architectures, parameters, and training algorithms, is retained and executed solely on the central server in \modelname. 
Besides, it is important to point out that, in \modelname, traditional model extraction attacks~\cite{zhang2024defense} that aim to steal models in a centralized paradigm may also be inapplicable and unaffordable, as these methods require a proportion of real user data while in \modelname each user can only access its own data.
Consequently, intellectual property remains uncompromised when adopting \modelname as a training paradigm.
Meanwhile, to protect user privacy, we introduce perturbations based on the exponential mechanism before users share sequences with the central server, making it challenging for the server to identify interacted items. 
It's worth noting that this scenario resembles traditional FedRecs, where the central server can infer trained items, comprising both interacted and negative items, but cannot accurately filter out interacted items related to user privacy. 
Consequently, our proposed \modelname considers the privacy needs of both service providers and users.

\subsubsection{Communication Efficiency Discussion}
The traditional parameter transmission-based federated sequential recommendation's communication costs are positively correlated with the model size.
Specifically, the one-time communication cost for a client $u_{i}$ to the central server can be represented as $\zeta \times \text{size}(\mathbf{M})$, where $\zeta$ is the efficiency factor. 
The size of $\mathbf{M}$ is typically substantial, consisting either of a high-dimensional item embedding table or a complex encoder, as indicated in Section~\ref{sec_basic_srm}. 
In contrast, for \modelname, the communication cost from the client to the central server primarily depends on the size of $\hat{\mathcal{D}}_{u_{i}}^{l}$, which essentially comprises $T_{u_{i}}$ integers.

\section{Experiments}\label{sec_experiments}
In this section, we conduct experiments to investigate the following research questions:
\begin{itemize}
  \item \textbf{RQ1.} How effective is our \modelname compared to centralized and conventional federated counterparts in recommendation performance?
  \item \textbf{RQ2.} How efficient is our \modelname compared to conventional federated counterparts in communication costs?
  \item \textbf{RQ3.} What are the impacts of the privacy-preserving client sequence uploading method?
  \item \textbf{RQ4.} What are the impacts of two contrastive learning auxiliary tasks in \modelname?
  \item \textbf{RQ5.} What are the impacts of the similarity-based knowledge downloading mechanism?
  \item \textbf{RQ6.} What is the influence of client model type?
\end{itemize}

\subsection{Datasets}
We employ three real-world datasets, Amazon Cell Phone, Amazon Baby~\cite{ni2019justifying}, and MIND~\cite{wu2020mind}, to evaluate the effectiveness of \modelname.
These datasets cover different recommendation domains, including electronics, baby products, and online news, originating from two famous platforms Amazon and Microsoft News.
The detailed statistics of each dataset are displayed in Table~\ref{tb_statistics}.
Amazon Cell Phone consists of $13,174$ users and $5,970$ cell phone-related products with $103,593$ reviews.
There are more than $160,000$ interaction recordings between $19,000$ users and $7,050$ baby cares on Amazon Baby.
For MIND, we randomly sample a subset from the original dataset due to the computation resource limitation, and the subset contains $6,260$ users, $14,505$ news, and $96,125$ reading histories.
We follow the most common data preprocessing procedure in sequential recommendation~\cite{jannach2017recurrent,kang2018self,yuan2023go} for all datasets.
All the presence of ratings or reviews is transformed to implicit feedback, i.e., $r=1$, and then, we sort them with their interacted timestamp to get the user interaction sequence.
Users who have less than five interactions are discarded and the maximum sequence length is set to $20$.
The last two items in a sequence are used for validation and test purposes respectively.

\begin{table}[!htbp]
  \centering
  \caption{Statistics of three datasets used in our experiments.}\label{tb_statistics}
  \begin{tabular}{l|cccc}
  \hline
  \textbf{Dataset}        & \textbf{Cell Phone} & \textbf{Baby} & \textbf{MIND}  \\ \hline
  \textbf{\#Users}        & 13,174      & 19,445        & 6,260           \\
  \textbf{\#Items}        & 5,970       & 7,050         & 14,505           \\
  \textbf{\#Interactions} & 103,593     & 160,792       & 96,125        \\
  \textbf{Avgerage Lengths} & 7.86      & 8.26          & 15.35         \\
  \textbf{Density} & 0.13\%             & 0.11\%        & 0.10\%      \\ \hline
  \end{tabular}
  \end{table}

\subsection{Evaluation Metrics}
We adopt two popular recommendation metrics~\cite{yuan2023federated,sankar2020groupim,yuan2024robust} Hit Ratio at rank 20 (HR@20) and Normalized Discounted Cumulative Gain at rank 20 (NDCG@20) to measure the model performance.
HR@20 evaluates the ratio of golden items included in the top-20 list, and NDCG@20 measures whether these items are ranked in the high position. 
We calculate the metrics scores for all items that have not interacted with users to avoid evaluation bias~\cite{krichene2020sampled}.

\subsection{Baselines}
We compare \modelname with several baselines including both centralized and federated sequential recommendation methods.

\textbf{Centralized Sequential Recommendation Baselines.}
We utilize GRU4Rec~\cite{jannach2017recurrent}, SASRec~\cite{kang2018self}, and MoRec~\cite{yuan2023go} as the centralized sequential recommendation baselines.
Note that we also leverage these three models as the base model in our \modelname.
Consequently, this comparison can directly showcase the performance gap between the centralized training paradigm and our federated training paradigms.

\textbf{Federated Sequential Recommendation Baselines.}
We select FMSS~\cite{lin2022generic} as our baseline from the existing federated sequential recommendation works as it is the only open-source framework that focuses on client-level federated sequential recommendation.
It designs fake marks and secret sharing with GRU4Rec to achieve distributed collaborative learning.

To make a more comprehensive comparison, we leverage the general federated learning framework discussed in Section~\ref{sec_general_fedrec} with the base recommendation model used in \modelname to form Fed-GRU4Rec and Fed-SASRec.
We do not implement Fed-MoRec since it requires clients to have impractical computation ability and suffer an unaffordable communication burden to train large language models.
By comparing with Fed-GRU4Rec and Fed-SASRec based on traditional federated learning framework, we can directly exhibit the advantages of our novel parameter transmission-free federated sequential framework.

\subsection{Hyperparameter Details} \label{sec_implementation_details}
In \modelname, we default to assigning SASRec as the client model, while the server models can be GRU4Rec, SASRec, or MoRec. 
In Section~\ref{sec_further_analysis}, we also present results utilizing GRU4Rec as client models. 
The size of the client model is set to be smaller than the server model due to the limitation of client training resources.
Specifically, when using GRU4Rec or SASRec as the client model, both the dimensions of item embeddings and hidden vectors are set to $8$, with a single neural network block layer. 
Conversely, when leveraging GRU4Rec and SASRec as server models, the item embedding and hidden vector sizes are $32$, and two layers of corresponding neural network blocks are stacked according to~\cite{kang2018self}. 
The MoRec on the server side is implemented based on the BERT-small version from the original paper due to the limitation of our computational resources. 
When implementing centralized and federated baselines, these models' sizes are consistent with the server version in \modelname for fair comparison.

The default privacy parameters $\epsilon$ and $\beta$ are set to $1.0$ and $0.5$ respectively, and their effects will be investigated in Section~\ref{sec_impact_privacy}. 
The contrastive learning controllers $\lambda_{pc}$ and $\lambda_{is}$ are both set to $0.01$, and their impacts are presented in Section~\ref{sec_impact_contrastive}. 
Both clients and the central server transfer only one sequence to each other every round for communication efficiency. 
The maximum global rounds are $20$.
Note that all clients will be trained in a global round.
Specifically, at the beginning of a global round, we first shuffle the client queue, and then, we traverse the queue with several subround by selecting $256$ clients each time to participate in the training process.
The local training epochs for clients and the central server are $5$ and $2$, respectively, following~\cite{yuan2023hide}.
For the server training, the data batch size is $1024$ while the batch size for the client model is set to equal its whole sequence numbers as it only has one private sequence combined with a few central server dispersed sequences.

\begin{table}[!htbp]
  \centering
  \caption{The recommendation performance of \modelname and baselines on three datasets. \modelnamenospace(X) represents that the central server utilizes model ``X'', meanwhile the clients utilize SASRec by default. The best performance of centralized recommendation is highlighted with underline, while the best performance of FedRecs is indicated by bold.}\label{tb_main}
  \begin{tabular}{l|cc|cc|cc}
    \hline
    \multicolumn{1}{c|}{\multirow{2}{*}{\textbf{Methods}}}                                                               & \multicolumn{2}{c|}{\textbf{Cell Phone}} & \multicolumn{2}{c|}{\textbf{Baby}}      & \multicolumn{2}{c}{\textbf{MIND}}       \\
    \multicolumn{1}{c|}{}                                                                                                & \textbf{HR@20}      & \textbf{NDCG@20}   & \textbf{HR@20}     & \textbf{NDCG@20}   & \textbf{HR@20}     & \textbf{NDCG@20}   \\ \hline
      \textbf{GRU4Rec}                                   & 0.0710           & 0.0299          & 0.0375          & 0.0141          & 0.1309   & 0.0581   \\
                                                \textbf{SASRec}                                    & 0.0869           & 0.0389          & 0.0386          & 0.0156          & \underline{0.1525}         & \underline{0.0700}        \\
                                                \textbf{MoRec}                                     & \underline{0.1098}     & \underline{0.0453}    & \underline{0.0561}    & \underline{0.0225}    & 0.0690          & 0.0286          \\ \hline
                                                \textbf{FMSS}                                      & 0.0541           & 0.0197          & 0.0253          & 0.0094          & 0.0744          & 0.0253          \\
        \textbf{Fed-GRU4Rec}                               & 0.0529           & 0.0193          & 0.0240          & 0.0091          & 0.0702          & 0.0243          \\
                                                \textbf{Fed-SaSRec}                                & 0.0520           & 0.0189          & 0.0304          & 0.0104          & 0.0661          & 0.0230          \\
                                                \textbf{\modelnamenospace(GRU4Rec)} & 0.0879           & 0.0400          & 0.0355          & 0.0145          & 0.1316                  & 0.0622                  \\
                                                \textbf{\modelnamenospace(SASRec)}  & 0.0973           & 0.0439          & 0.0430          & 0.0176          & \textbf{0.1380} & \textbf{0.0626} \\
                                                \textbf{\modelnamenospace(MoRec)}   & \textbf{0.1260}  & \textbf{0.0550} & \textbf{0.0541} & \textbf{0.0208} & 0.0840          & 0.0359          \\ \hline
    \end{tabular}
  \end{table}

\subsection{Effectiveness of \modelname (RQ1)}\label{sec_main_experiment}
The comparison of recommendation performance between \modelname and the baselines is presented in Table~\ref{tb_main}. \modelnamenospace(X) indicates that the central server utilizes model X while clients' models are simpler versions of SASRec as described in Section~\ref{sec_implementation_details}.

Generally, traditional federated sequential recommendation methods fail to achieve satisfactory performance compared to centralized sequential recommendation and our \modelname in all cases. This discrepancy may stem from the fact that in the sequential recommendation, each client contains only one interaction sequence, while the sequential recommendation model is more complex than collaborative filtering models. Consequently, the local gradients calculated on a single sequence often deviate significantly from the optimal point, resulting in suboptimal performance.

Furthermore, our \modelname achieves comparable or even superior performance compared to the corresponding centralized model versions.
Specifically, on Cell Phone dataset, all \modelname models achieve better performance than their centralized counterparts, and \modelnamenospace(MoRec) obtains the best performance with $0.12$ HR@20 and $0.05$ NDCG@20 scores.
On the Baby and MIND datasets, our \modelname models' performances slightly lag behind their centralized versions, but they outperform other FedSeqRecs by significant margins.

Additionally, by comparing different model types across datasets, we observe that in both centralized paradigms and our \modelname, the more complex the sequential models are, the better performance they achieve on the Cell Phone and Baby datasets. However, the trend is reversed on the MIND dataset. This phenomenon can be attributed to the data statistics shown in Table~\ref{tb_statistics}. The average interaction counts of items in MIND are much lower than those in the other two datasets, which cannot support more complex model training.

\begin{table}[!htbp]
  \centering
  \caption{The comparison of average communication costs per client for one round. The costs for \modelnamenospace(GRU4Rec), \modelnamenospace(SASRec), and \modelnamenospace(MoRec) are the same, thus we report them as \modelname to avoid repetition. The most efficient costs are indicated by bold.}\label{tb_communication}
  \begin{tabular}{lccc}
  \hline
  \textbf{Methods} & \textbf{Cell Phone} & \textbf{Baby} & \textbf{MIND} \\ \hline
  \textbf{FMSS}          & 6.2MB                & 7.3MB             & 14.8MB           \\
  \textbf{Fed-GRU4Rec}   & 3.0MB                & 3.6MB             & 14.4MB          \\
  \textbf{Fed-SASRec}    & 1.6MB                & 1.8MB             & 3.8MB          \\ 
  \textbf{Fed-MoRec}(Est. cost)    & 234MB                & 234MB             & 234MB          \\ 
  \textbf{\modelname}    & \textbf{1.2}KB                & \textbf{1.3}KB              & \textbf{2.4}KB           \\ \hline
  \end{tabular}
  \end{table}
\subsection{Communication Efficiency of \modelname (RQ2)}\label{sec_impact_communication}
Aside from the final model performance, the communication burden is the other important feature when evaluating a federated recommendation framework.
Table~\ref{tb_communication} illustrates the communication costs of a one-time interaction between a single client and the central server. Since \modelname's communication burden is decoupled from models, the costs for \modelnamenospace(GRU4Rec), \modelnamenospace(SASRec), and \modelnamenospace(MoRec) are the same. According to the results, the communication costs of traditional FedSeqRecs are positively correlated with model sizes. Therefore, for ID-based sequential recommendation models (e.g., FMSS, Fed-GRU4Rec, and Fed-SASRec), as the number of items increases (i.e., from Cell Phone to MIND), the communication costs also increase. On the other hand, the communication costs for ID-free sequential recommendation models (e.g., MoRec) are unrelated to item numbers, however, they remain unaffordable. It is worth noting that the costs of Fed-MoRec are only estimated since we lack sufficient computational power to train such a large number of MoRecs for numerous clients.

The communication costs of our \modelname are solely related to the length of user interaction logs. 
In the three datasets used in this paper, the average cost per client-server communication is no more than $3$KB, as the average user sequence lengths are below $16$. 
In real-world applications, the average user interaction sequence is unlikely to be excessively long due to data sparsity.
Additionally, for a few very active users possessing long interaction sequences, the system will typically set a maximum sequence length to ensure the model learns from the most recently interacted items. 
Therefore, the communication costs of \modelname remain consistently lightweight.

\begin{figure}[!ht]
  \centering
  \subfloat[]{\includegraphics[width=0.8\textwidth]{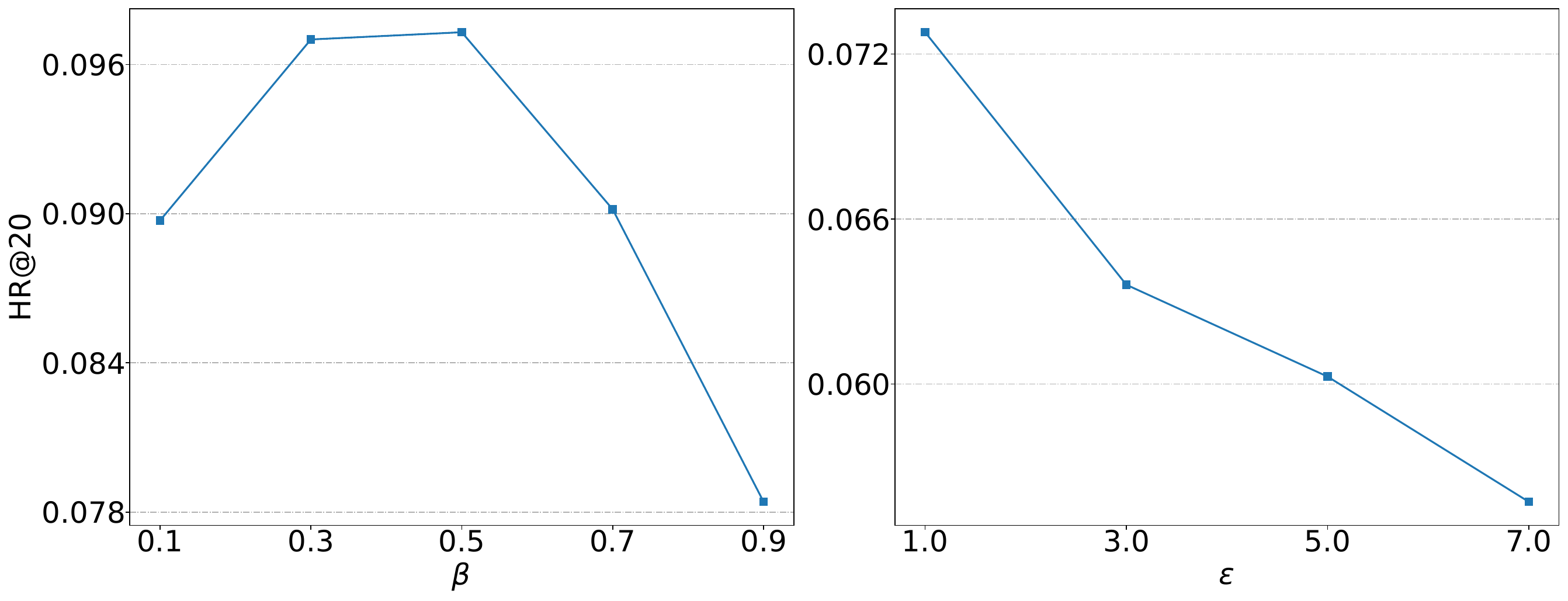}\label{fig_sas_privacy}}
  \hfil
  \subfloat[]{\includegraphics[width=0.8\textwidth]{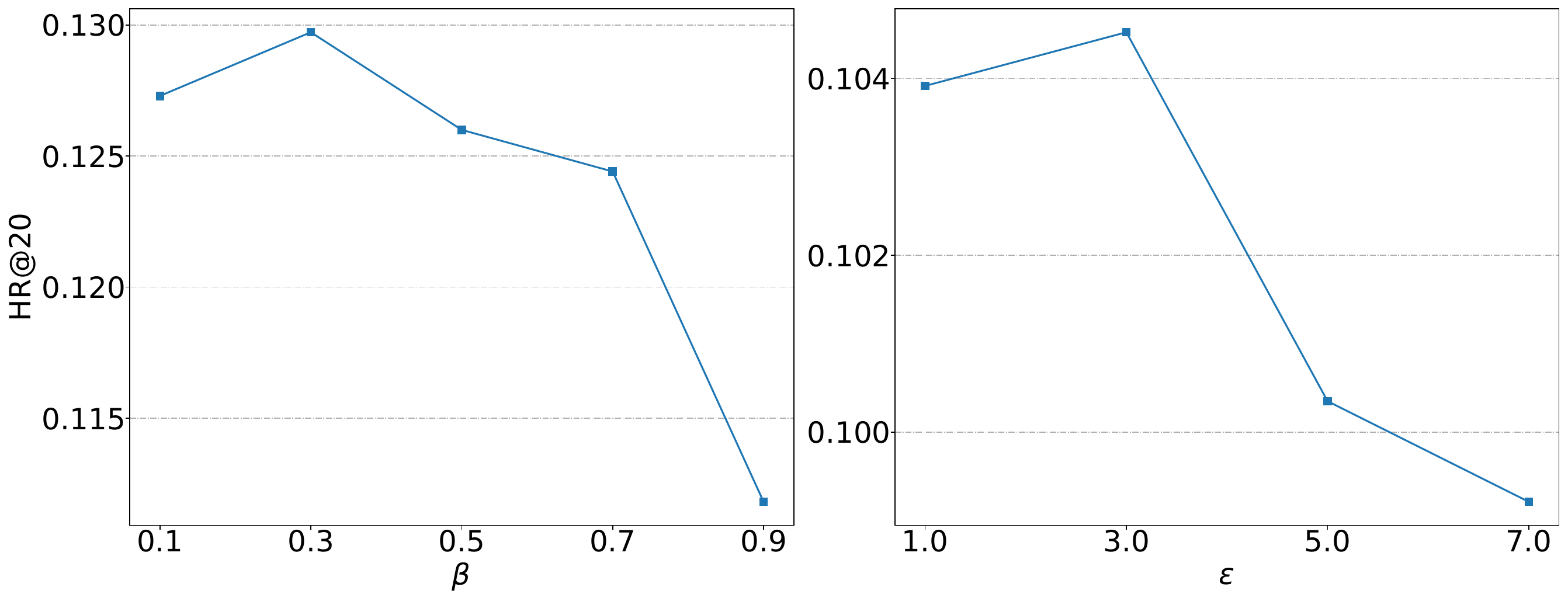}\label{fig_morec_privacy}}
  \caption{The impact of privacy parameters $\beta$ and $\epsilon$ for (a) \modelnamenospace(SASRec) and (b) \modelnamenospace(MoRec) on Cell Phone dataset. Similar trends can be observed in the other two datasets. Note that when we investigate one hyperparameter, the other is set to the default value mentioned in Section~\ref{sec_implementation_details}. That is to say, $\epsilon=1.0$ when $\beta$ changes and $\beta=0.5$ when $\epsilon$ be modified.}\label{fig_privacy_hyperparameter}
\end{figure}

\begin{table}[!ht]
  \centering
  \caption{The comparison of recommendation performance of \modelname on three datasets when using similarity knowledge sharing (+SK) or using randomly selected knowledge sharing (-SK).}\label{tb_sks}
  \begin{tabular}{ll|cc|cc}
    \hline
                                                              &                  & \multicolumn{2}{c|}{\textbf{\modelnamenospace(SASRec)}} & \multicolumn{2}{c}{\textbf{\modelnamenospace(MoRec)}} \\ \cline{3-6} 
                                                              &                  & \multicolumn{1}{c}{\textbf{+SK}}  & \multicolumn{1}{c|}{\textbf{-SK}}  & \multicolumn{1}{c}{\textbf{+SK}}  & \multicolumn{1}{c}{\textbf{-SK}} \\ \hline
    \multicolumn{1}{l|}{\multirow{2}{*}{\textbf{Cell Phone}}} & \textbf{HR@20}   & \textbf{0.0973}                   & 0.0844                            & \textbf{0.1260}                   & 0.1169                           \\
    \multicolumn{1}{l|}{}                                     & \textbf{NDCG@20} & \textbf{0.0439}                   & 0.0355                            & \textbf{0.0550}                   & 0.0497                           \\ \hline
    \multicolumn{1}{l|}{\multirow{2}{*}{\textbf{Baby}}}       & \textbf{HR@20}   & \textbf{0.0430}                   & 0.0358                            & \textbf{0.0541}                   & 0.0510                           \\
    \multicolumn{1}{l|}{}                                     & \textbf{NDCG@20} & \textbf{0.0176}                   & 0.0133                            & \textbf{0.0208}                   & 0.0199                           \\ \hline
    \multicolumn{1}{l|}{\multirow{2}{*}{\textbf{MIND}}}       & \textbf{HR@20}   & \textbf{0.1380}                   & 0.1062                            & \textbf{0.0840}                   & 0.0811                           \\
    \multicolumn{1}{l|}{}                                     & \textbf{NDCG@20} & \textbf{0.0626}                   & 0.0446                            & \textbf{0.0359}                   & 0.0346                           \\ \hline
    \end{tabular}
  \end{table}

\subsection{The Impact of Privacy-preserving Mechanism (RQ3)}\label{sec_impact_privacy}
In this section, we evaluate the influence of two privacy-preserving parameters, $\beta$ and $\epsilon$, on model performance. 
$\beta$ controls the ratio of items in a sequence that will undergo exponential mechanism generation, while $\epsilon$ determines the randomness of the exponential mechanism-based generation.
Fig.\ref{fig_privacy_hyperparameter} illustrates the performance changes with these two hyperparameters on the Cell Phone dataset with \modelnamenospace(SASRec) (Fig.\ref{fig_sas_privacy}) and \modelnamenospace(MoRec) (Fig.~\ref{fig_morec_privacy}). 
We apply \modelname with these two models as they represent ID-based and ID-free sequential recommendations and yielded satisfactory results in our main experiments. 
Similar trends can be observed in the other two datasets.

Generally, as the replacement ratio $\beta$ increases, the model performance initially improves and then rapidly declines. Specifically, the model performance peaks at $\beta=0.5$ for \modelnamenospace(SASRec) and $\beta=0.3$ for \modelnamenospace(MoRec), after which it decreases sharply with further increases in $\beta$.
This occurs because an appropriate replacement ratio $\beta$ can serve as an augmentation method for our contrastive denoising mechanism described in Section~\ref{sec_server_advanced_training}, thereby enhancing model performance.
However, when $\beta$ becomes too large, the sequence becomes random and loses most of its original semantics, rendering it difficult for the central server model to learn meaningful patterns from these random sequences.

Normally, increasing the privacy budget leads to better performance as the expected results (i.e., the items with the largest prediction scores) become more determined to be selected.
However, in our experiments, we found a negative correlation between $\epsilon$ and model performance.
This arises from the insufficient training of local sequential models in the initial few rounds, causing their expected items to be incorrect and deviating from the optimization process.
Specifically, the differences in prediction scores for some potential items are not very apparent at the beginning of a few rounds, and the model is still learning about them with less confidence. 
A large $\epsilon$ value just steepens the prediction distribution and increases the likelihood of selecting the currently largest prediction items.

\begin{figure}[!htbp]
  \centering
  \subfloat[]{\includegraphics[width=0.8\textwidth]{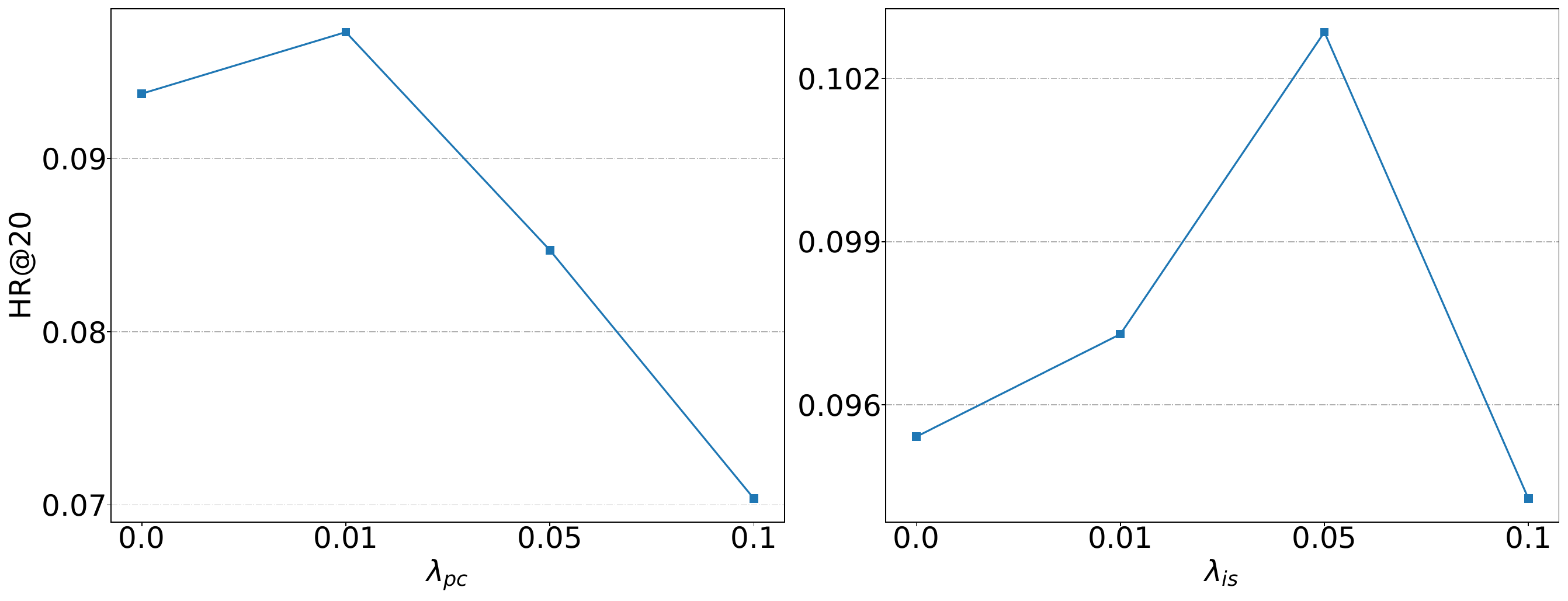}\label{fig_sas_contrastive}}
  \hfil
  \subfloat[]{\includegraphics[width=0.8\textwidth]{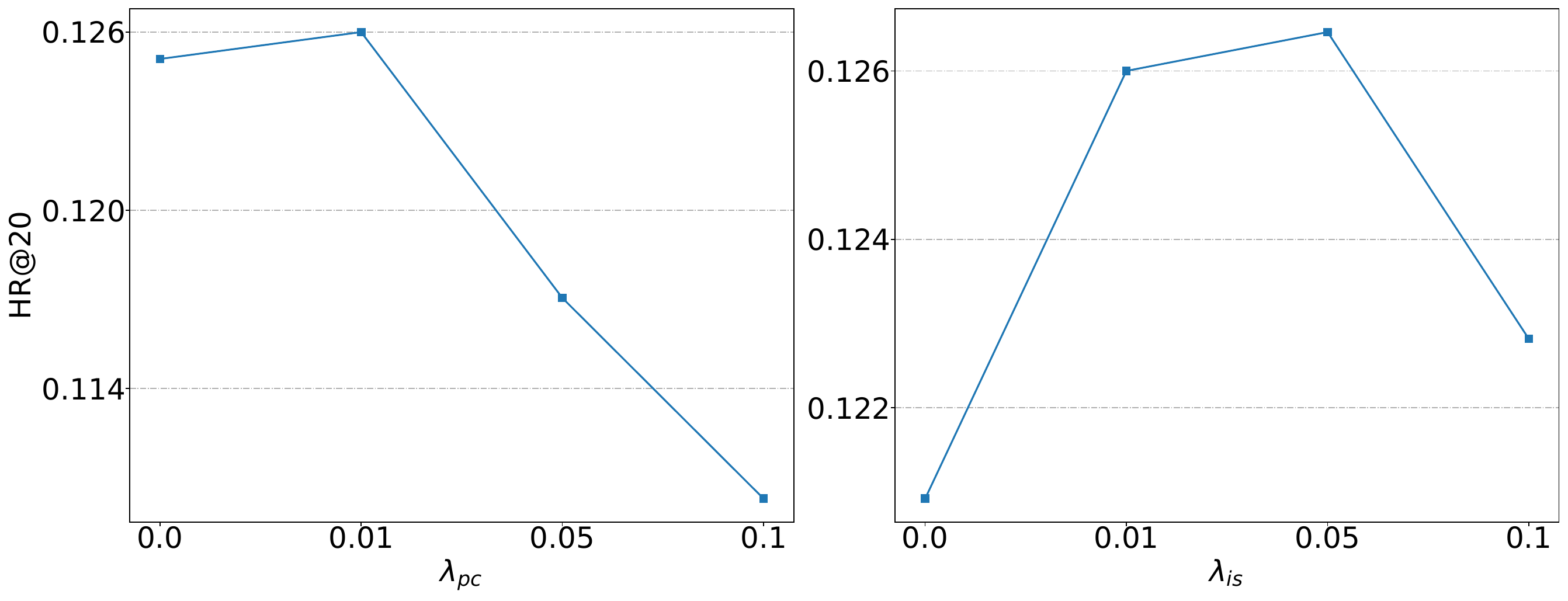}\label{fig_morec_contrastive}}
  \caption{The impact of contrastive factors $\lambda_{pc}$ and $\lambda_{is}$ for (a) \modelnamenospace(SASRec) and (b) \modelnamenospace(MoRec) on Cell Phone dataset. Similar trends can be observed in the other two datasets. When we investigate one factor, another factor is keeping the default value, i.e., $\lambda_{pc/is}=0.01$.}\label{fig_contrastive_hyperparameter}
\end{figure}

\subsection{The Impact of Contrastive Denoising Methods (RQ4)}\label{sec_impact_contrastive}
To enhance our server model's ability to learn from noisy user-uploaded sequences, we introduce preference consistency contrastive learning $\mathcal{L}^{pc}$ and intention similarity contrastive learning $\mathcal{L}^{is}$ as denoising auxiliary tasks. $\lambda_{pc}$ and $\lambda_{is}$ are two factors controlling the strengths of these tasks' contributions.

Fig.~\ref{fig_contrastive_hyperparameter} illustrates the performance trends of \modelnamenospace(SASRec) and \modelnamenospace(MoRec) with increasing $\lambda_{pc}$ and $\lambda_{is}$ on the Cell Phone dataset, with similar trends observed on the other two datasets.
When $\lambda_{pc}<0.01$, increasing the values of this factor positively influences the final model performance, confirming the effectiveness of preference consistency learning. However, with larger $\lambda_{pc}$ (i.e., $\lambda_{pc}>0.01$), the auxiliary task $\mathcal{L}^{pc}$ becomes overwhelming and significantly deteriorates the model performance.

The trend of $\lambda_{is}$ is very similar to $\lambda_{pc}$. 
When an appropriate value of $\lambda_{is}$ is chosen, the $\mathcal{L}^{is}$ task positively contributes to the model training compared to $\lambda_{is}=0$ where $\mathcal{L}^{is}$ is totally removed.
However, excessively large $\lambda_{is}$ values lead to a drop in performance.
In summary, the effectiveness of both $\mathcal{L}^{pc}$ and $\mathcal{L}^{is}$ has been demonstrated in the results.

  \begin{table}[!ht]
    \centering
    \caption{The performance of different model combinations for clients and the server.}\label{tb_model_type}
    \begin{tabular}{l|l|cc|cc|cc}
      \hline
                                        & \textbf{}        & \multicolumn{2}{c|}{\textbf{Cell Phone}} & \multicolumn{2}{c|}{\textbf{Baby}} & \multicolumn{2}{c}{\textbf{MIND}}  \\ \hline
      \textbf{Server$\downarrow$}                   & \textbf{Client$\rightarrow$}  & \textbf{GRU4Rec}    & \textbf{SASRec}    & \textbf{GRU4Rec} & \textbf{SASRec} & \textbf{GRU4Rec} & \textbf{SASRec} \\ \hline
      \multirow{2}{*}{\textbf{GRU4Rec}} & \textbf{HR@20}   & 0.0846              & 0.0879             & 0.0342           & 0.0355          & 0.1223           & 0.1316          \\
                                        & \textbf{NDCG@20} & 0.0383              & 0.0400             & 0.0131           & 0.0145          & 0.0559           & 0.0622          \\
      \multirow{2}{*}{\textbf{SASRec}}  & \textbf{HR@20}   & 0.0975              & 0.0973             & 0.0398           & 0.0430          & \textbf{0.1413}  & 0.1380          \\
                                        & \textbf{NDCG@20} & 0.0440              & 0.0439             & 0.0161           & 0.0176          & \textbf{0.0631}  & 0.0626          \\
      \multirow{2}{*}{\textbf{MoRec}}   & \textbf{HR@20}   & 0.1194              & \textbf{0.1260}    & 0.0507           & \textbf{0.0541} & 0.0758           & 0.0840          \\
                                        & \textbf{NDCG@20} & 0.0504              & \textbf{0.0550}    & 0.0185           & \textbf{0.0208} & 0.0302           & 0.0359          \\ \hline
      \end{tabular}
    \end{table}

\subsection{The Impact of Similarity Knowledge Sharing (RQ5)}\label{sec_impact_of_simi_share}
Facilitating the training of client models can ultimately enhance the performance of the server model, as the latter learns from the sequences generated by the former. With this in mind, \modelname employs a similarity-based knowledge-sharing approach by disseminating sequences that exhibit significant semantic similarity to local clients. In this section, we aim to validate the effectiveness of this mechanism.

We compare the similarity knowledge sharing (+SK) with a randomly selected knowledge sharing method (-SK) in Table~\ref{tb_sks} across three datasets using \modelnamenospace(SASRec) and \modelnamenospace(MoRec). Clearly, on all three datasets, models trained with randomly selected knowledge-sharing mechanisms fail to outperform our similarity-based knowledge-sharing method. 
For instance, on the Cell Phone dataset, the HR@20 scores of \modelnamenospace(SASRec) drop from $0.097$ to $0.084$ when transitioning from similarity sharing to random sharing, while those of \modelnamenospace(MoRec) decrease from $0.126$ to $0.116$.
Overall, these results underscore the effectiveness of similarity knowledge sharing.

\subsection{Analysis on Client Model Type (RQ6)}\label{sec_further_analysis}
Essentially, \modelname implements model privacy protection by achieving server and client model heterogeneity. 
In the main experiments, we default to setting the client model as SASRec and explore various models on the central server side. 
In this section, we further analyze all model combinations for client and server models across all three datasets and present the results in Table~\ref{tb_model_type}. Note that since clients' computational power usually cannot support the training of large language model-based recommender systems, we do not consider MoRec as a client model.

Firstly, by comparing different server models within the same client model type (vertical comparison in Table~\ref{tb_model_type}), we observe that the more advanced the server models are, the better final performance they achieve on the Cell Phone and Baby datasets, since these two datasets have relatively sufficient training sources.
However, when the average training corpus for each item is relatively limited, such as in the MIND dataset, employing complex models leads to worse performance. Additionally, the comparison between different client model types within the same server models (horizontal comparison in Table~\ref{tb_model_type}) indicates that using SASRec as the client model achieves better performance in most cases.

\section{Conclusion and Future Work} \label{sec_conclusion}
In this paper, we extend our previous parameter transmission-free federated recommendation framework~\cite{yuan2023hide} to the sequential recommendation task, namely \modelname, to address the issues of protecting model intellectual property and reducing heavy communication burdens. 
In \modelname, client models and central server models are heterogeneous, and they achieve collaborative learning by sharing generated sequences. 
To protect user data privacy, \modelname is equipped with an exponential mechanism-based method to add noise to the original sequences. 
Furthermore, we design two contrastive denoising tasks to compel the server model to learn high-level representations. 
A similarity-based knowledge-sharing approach is employed to transfer the server model's knowledge to the client side. 
Extensive experiments with both traditional ID-based sequential models and current large language model-based ID-free sequential recommender systems on three recommendation datasets demonstrate the superiority of \modelname.

As a new design of a federated sequential recommendation framework, this paper represents an initial step, and there are many research aspects that can be further explored in the future. For example, since model parameters are not used as knowledge carriers, the models can not only be heterogeneous between clients and the central server, as explored in this paper, but also among clients themselves. Therefore, it is promising to investigate the optimal client model deployment strategy considering clients' local resources (e.g., computational resources, data resources, etc.). For clients with sufficient training resources, a relatively strong model can be deployed. However, for ``poor'' clients, a smaller model may be a good choice to reduce their training burden.

\begin{acks}
  This work is supported by the Australian Research Council under the streams of Future Fellowship (Grant No. FT210100624) and the Discovery Project (Grant No. DP240101108). 
\end{acks}

\bibliographystyle{ACM-Reference-Format}
\bibliography{sample-base}










\end{document}